\documentclass[mathleft]{an}
\usepackage{times}
\usepackage{graphicx}
\usepackage{url}
\usepackage{bm}
\graphicspath{{./fig/}{./png/}}

\newcommand{\EQ}{\begin{equation}}
\newcommand{\EN}{\end{equation}}
\newcommand{\EQA}{\begin{eqnarray}}
\newcommand{\ENA}{\end{eqnarray}}

\newcommand{\Eq}[1]{Eq.~(\ref{#1})}

\newcommand{\Fig}[1]{Fig.~\ref{#1}}

\newcommand{\Tab}[1]{Table~\ref{#1}}

\newcommand{\bra}[1]{\langle #1\rangle}

\newcommand{\meanBB}{\overline{\vec{B}}}

\newcommand{\meanUU}{\overline{\vec{U}}}

\newcommand{\ww}{\mbox{\boldmath $w$} {}}
\newcommand{\WW}{\mbox{\boldmath $W$} {}}

\newcommand{\UU}{{\vec{U}}}
\newcommand{\uu}{{\vec{u}}}
\newcommand{\BB}{{\vec{B}}}
\newcommand{\JJ}{{\vec{J}}}

\newcommand{\AAA}{{\vec{A}}}

\newcommand{\ff}{\mbox{\boldmath $f$} {}}

\newcommand{\nab}{\mbox{\boldmath $\nabla$} {}}

\newcommand{\SSSS}{\mbox{\boldmath ${\sf S}$} {}}

\newcommand{\dd}{{\rm d} {}}
\newcommand{\const}{{\rm const}  {}}

\def\Pm{\mbox{\rm Pr}_{\it M}}
\def\Rm{\mbox{\rm Re}_{\it M}}
\def\Rmc{\mbox{\rm Re}_{\it M}^{\rm crit}}
\def\Rey{\mbox{\rm Re}}

\def\cs{c_{\rm s}}
\def\kf{k_{\rm f}}
\def\urms{u_{\rm rms}}
\def\uoneD{u_{\rm 1D}}
\def\epsf{\epsilon_{\rm f}}
\def\epsK{\epsilon_{\it K}}
\def\epsM{\epsilon_{\it M}}
\def\epsT{\epsilon_{\it T}}
\def\etat{\eta_{\rm t}}
\def\etaT{\eta_{\rm T}}

\def\half{{\textstyle{1\over2}}}

\def\onethird{{\textstyle{1\over3}}}

\newcommand{\yapj}[3]{: #1, {ApJ} {#2}, #3}

\newcommand{\ypf}[3]{: #1, {PhFl} {#2}, #3}

\newcommand{\yjetp}[3]{: #1, {Sov. Phys. JETP} {#2}, #3}

\newcommand{\yaraa}[3]{ #1, {ARA\&A,} {#2}, #3}

\newcommand{\yprl}[3]{: #1, {Phys Rev Lett} {#2}, #3}
\newcommand{\ypre}[3]{: #1, {Phys Rev E} {#2}, #3}

\newcommand{\ymn}[3]{: #1, {MNRAS} {#2}, #3}

\newcommand{\yjour}[4]{: #1, {#2} {#3}, #4}

\Pagespan{725}{}
\Yearpublication{2011}
\Yearsubmission{2010}
\Month{1}
\Volume{332}
\Issue{1}
\DOI{10.1002/asna.200811027}

\begin{document}

\title{Dissipation in dynamos at low and high magnetic Prandtl numbers}
\authorrunning{A. Brandenburg}
\author{A. Brandenburg\thanks{Corresponding author: brandenb@nordita.org   }}
\institute{
NORDITA, AlbaNova University Center, Roslagstullsbacken 23, SE-10691 Stockholm, Sweden; and\\
Department of Astronomy, Stockholm University, SE 10691 Stockholm, Sweden
}

\received{2010 Oct 22}  \accepted{2010 Nov 18}
\publonline{2010 Dec 30}


\abstract{%
Using simulations of helically driven turbulence, it is shown that the
ratio of kinetic to magnetic energy dissipation scales with the magnetic
Prandtl number in power law fashion with an exponent of approximately 0.6.
Over six orders of magnitude in the magnetic Prandtl number
the magnetic field is found to be sustained by large-scale
dynamo action of alpha-squared type.
This work extends a similar finding for small magnetic Prandtl numbers
to the regime of large magnetic Prandtl numbers.
At large magnetic Prandtl numbers, most of the energy is dissipated
viscously, lowering thus the amount of magnetic energy dissipation, which
means that simulations can be performed at magnetic Reynolds numbers
that are large compared to the usual limits imposed by a given resolution.
This is analogous to an earlier finding that at small magnetic Prandtl
numbers, most of the energy is dissipated resistively, lowering the
amount of kinetic energy dissipation, so simulations can then be performed
at much larger fluid Reynolds numbers than otherwise.
The decrease in magnetic energy dissipation at large magnetic Prandtl
numbers is discussed in the context of underluminous accretion found in
some quasars.
}

\maketitle

\section{Introduction}

The magnetic fields in astrophysical bodies often have a pronounced
large-scale component that is associated with large-scale dynamo action.
Examples are the cyclic magnetic fields in late-type stars such as the Sun
and the magnetic spirals in many galaxies, including even irregular
galaxies; see Beck et al.\ (1996) for a review.
In addition, all observed magnetic fields also have a significant
small-scale component that may either be the result of turbulent
motions distorting the large-scale field, or, alternatively, it
could be the result of what is known as small-scale dynamo action
(Cattaneo 1999).

Much of our knowledge about large-scale and small-scale dynamos
has come from numerical simulations; see Brandenburg \& Subramanian
(2005) for a review.
It is clear that, in order for simulations to approach an astrophysically
interesting regime, one wants to make both the magnetic diffusivity and
the kinematic viscosity as small as possible.
This means that the magnetic and fluid Reynolds numbers should be as
large as possible for a given numerical resolution, $N^3$.
The relevant criterion for sufficient numerical resolution
is that the kinetic and magnetic energy spectra
should develop an exponentially decaying dissipative subrange at a
wavenumber that is at least a factor of 10 below the Nyquist frequency,
$k_{\rm Ny}=\pi N/L$.
In practice, for example, with a simulation at a resolution of $512^3$
mesh points, one can hardly exceed values of the magnetic and fluid
Reynolds number of about 500--700 (e.g., Brandenburg 2009).
However, as will be discussed in more detail in this paper, this empirical
constraint on the resolution really only applies if the ratio of magnetic
and fluid Reynolds numbers is about unity.
This ratio is also referred to as the magnetic Prandtl number, $\Pm$,
and there is hardly any system where this number is unity.
In galaxies and galaxy clusters this number tends to be very large,
while in stars and stellar accretion discs it is quite small.
Also liquid metals used in laboratory experiments have small $\Pm$.
Therefore, much of what has been learnt from numerical simulations at
$\Pm\approx1$ has to be re-examined in cases of low and high values
of $\Pm$.

The purpose of this paper is to focus on the relative importance of
viscous and ohmic dissipation rates at different values of $\Pm$.
Often, viscous and ohmic dissipation are only treated ``numerically''
by making sure the code is stable.
In such cases, viscosity and magnetic diffusivity are usually not even
stated explicitly in the equations, suggesting that these terms are
negligible and not important.
This is of course not the case, as can be illustrated by considering
the case of quasars that belong to the most luminous objects in the sky.
The discovery of the first quasar, 3C 273, is nicely explained by
Rhodes (1978) in a popular magazine.
Indeed, 3C 273, has about $2\times10^{12}$ times the luminosity
of the Sun and is indeed the brightest one in the sky.
This quasar would not shine at all if it was not for the effect of microphysical
viscosity that leads to viscous dissipation.
But how important is viscous dissipation compared with ohmic dissipation?
In order to address this problem we need to understand the effects of
both viscosity and magnetic diffusivity in a turbulent system where the
magnetic field is self-sustained by dynamo action.
In this paper we review briefly some recent work on dynamos in the regime
of small $\Pm$ and turn then to the investigation of large $\Pm$.

\section{Small magnetic Prandtl number dynamos}

In the last 6 years the issue of low magnetic Prandtl numbers,
$\Pm=\nu/\eta$, has become a frequently discussed topic in the dynamo
community.
This is the regime where the magnetic diffusivity $\eta$ is large
compared with the kinematic viscosity $\nu$.
Already over a decade ago, Rogachevskii \& Kleeorin (1997) noticed that
for small-scale dynamos the critical value of the magnetic Reynolds
number, $\Rm$, for the onset of dynamo action should rise from a value
around 35 at $\Pm=1$ to values around 400 for small values of $\Pm$.
Here, $\Rm=\urms/\eta\kf$ is defined with respect to the wavenumber
$\kf$ of the energy-carrying eddies and the rms velocity, $\urms$.
However, the result of Rogachevskii \& Kleeorin was not widely recognized
at the time.
In 2004, simulation began to address this point systematically.
Simulations of Schekochihin et al.\ (2004) and Haugen et al.\ (2004)
provided clear indications that $\Rmc$ rises, and the results of
Schekochihin et al.\ (2005) might have even suggested that the critical
value of $\Rm$ for small-scale dynamo action might have become infinite
for $\Pm\approx0.1$.

Meanwhile, Boldyrev \& Cattaneo (2004) provided an attractive framework for
understanding this behavior.
Given that the energy spectrum of the small-scale dynamo peaks
at the resistive scale, which is the smallest possible scale at
which the motions can still overcome resistive damping, one must
ask what are the properties of the flow at this scale.

In the original scenario of Kazantsev (1968), the small-scale dynamo
works through a velocity field that is random, but essentially
laminar and of large scale.
In a simulation this can be realized by choosing a large magnetic
Prandtl number, so the magnetic Reynolds number is much larger than
the fluid Reynolds number.
However, subsequent studies show that small-scale dynamo action
can also occur for $\Pm$ of order unity.
Both for $\Pm=1$ and for $\Pm\gg1$ one finds that the spectral
magnetic energy increases with wavenumber proportional to $k^{3/2}$.

A qualitatively new feature emerges when $\Pm$ is small.
In that case the wavenumber corresponding to the resistive scale decreases
and lies in the inertial range of the turbulence.
This property is crucial because in the inertial range the velocity field
is ``rough'', i.e.\ over a spatial interval $\delta x$ the velocity
difference $\delta u=u(x+\delta x)-u(x)$ scales like
$\delta u\sim\delta x^\zeta$ where $\zeta<1$.
Thus, the finite difference quotient of the velocity, $\delta u/\delta x$,
diverges with decreasing $\delta x$, provided $\delta x$ is still bigger
than the viscous cutoff scale.
According to Boldyrev \& Cattaneo (2004), the critical magnetic Reynolds
number increases with increasing roughness.

In all situations that have been simulated, the wavenumber range of the
spectra has been too limited so that they are affected by cutoff effects
both at large and small scales.
In particular, only in simulations beyond $1024^3$ meshpoints the spectra
are shallower than $k^{-5/3}$.
This is referred to as the bottleneck effect and is believed to be a
physical effect (Falkovich 1994, Dobler et al.\ 2003, Frisch et al.\ 2008).
One reason, however, why it is not usually seen in wind tunnel or
atmospheric boundary layer turbulence is the fact that one measures
in these cases only one-dimensional spectra.
In order to obtain three-dimensional spectra, one has to differentiate
those data, i.e.\ (Dobler et al.\ 2003)
\EQ
E_{\rm3D}=-\dd E_{\rm1D}/\dd\ln k.
\EN
Accepting thus the physical reality of the bottleneck effect,
it becomes plausible that the critical magnetic Reynolds number
for the onset of small-scale dynamo action reaches a maximum
around $\Pm=0.1$, and that it decreases somewhat for smaller
values of $\Pm$.
This is indeed what the simulations of Iskakov et al.\ (2007)
suggest.

Let us now switch to large-scale dynamos.
Their excitation conditions are characterized by the dynamo number
which, for helical turbulence and in the absence of shear, is just
\EQ
C_\alpha={\alpha\over\etaT k_1}\approx\epsf\iota{\kf\over k_1}.
\label{Calp}
\EN
Here, $k_1=2\pi/L$ is the minimal wavenumber in the domain of size $L$
and we have inserted standard approximations for the $\alpha$ effect,
$\alpha=\onethird\tau\overline{\ww\cdot\uu}$, and the turbulent
magnetic diffusivity, $\etat=\onethird\tau\overline{\uu^2}$.
Here, $\uu=\UU-\meanUU$ is the fluctuating velocity, i.e.\ the
difference between the actual velocity $\UU$ and the mean velocity
$\meanUU$, $\tau\approx(\urms\kf)^{-1}$ is the turnover time,
$\ww=\nab\times\uu$ is the fluctuating vorticity,
$\epsf=\overline{\ww\cdot\uu}/\kf\overline{\uu^2}$ is a measure for
the relative helicity, and $\iota=1+3/\Rm$ is a correction factor
of order unity for sufficiently large values of $\Rm$.
It turns out that in all cases the spectra of magnetic energy are
at the largest scale approximately independent of $\Rm$ for $\Pm$
between 1 and $10^{-3}$.
This was shown in Brandenburg (2009) and will here be extended to
$10\leq\Pm\leq10^3$.

At larger wavenumbers there is a striking difference in the magnetic
energy spectra between $\Pm=1$ and $\ll1$ in that the resistive cutoff
wavenumber moves toward smaller values.
At the same time, the kinetic energy spectrum becomes progressively
steeper, leaving less kinetic energy to dissipate.
This has two important consequences.
First of all, the fractional kinetic energy dissipation decreases
with decreasing $\Pm$ proportional to $\Pm^{1/2}$ (Brandenburg 2009).
On the other hand, the decrease of $\epsilon_{\rm K}$ implies that the
demand for numerical resolution becomes less stringent.
This, in turn, means that one can increase the value of $\Rey$ beyond
the normally established empirical limits.
An important goal of the present paper is the demonstration that the
same is also true in the opposite limit of $\Pm\gg1$.

\section{The model}

Our model is similar to that presented in Brandenburg (2001, 2009), where
we solve the hydromagnetic equations for velocity $\UU$, logarithmic
density $\ln\rho$, and magnetic vector potential $\AAA$ for an isothermal gas
in the presence of an externally imposed helical forcing function $\ff$,
\EQ
{\partial\UU\over\partial t}=-\UU\cdot\nab\UU-\cs^2\nab\ln\rho
+\ff+(\JJ\times\BB+\nab\cdot2\rho\nu\SSSS)/\!\rho,\,
\label{dUU}
\EN
\EQ
{\partial\ln\rho\over\partial t}=-\UU\cdot\nab\ln\rho-\nab\cdot\UU,
\label{dlnrho}
\EN
\EQ
{\partial\AAA\over\partial t}=\UU\times\BB-\mu_0\eta\JJ.
\label{indEq}
\EN
Here, $\BB=\nab\times\AAA$ is the magnetic field, $\JJ=\nab\times\BB/\mu_0$
is the current density, $\mu_0$ is the vacuum permeability,
$\cs$ is the isothermal speed of sound, and
${\sf S}_{ij}={1\over2}(U_{i,j}+U_{j,i})-{1\over3}\delta_{ij}\nab\cdot\UU$
is the traceless rate of strain tensor.
We consider a triply periodic domain of size $L^3$, so the smallest
wavenumber in the domain is $k_1=2\pi/L$.
The forcing function consists of eigenfunctions of the curl operator
with positive eigenvalues and is therefore fully helical with
$\ff\cdot\nab\times\ff=k\ff^2$, where $3.5\leq k/k_1\leq4.5$ is the
chosen wavenumber interval of the forcing function, whose average value
is referred to as $\kf\approx4\,k_1$.
The amplitude of $\ff$ is such that the Mach number is $\urms/\cs\approx0.1$,
so compressive effects are negligible (Dobler et al.\ 2003).
As in Brandenburg (2009), we choose as
initial conditions a Beltrami field of low amplitude.
The initial velocity is zero and the initial density is uniform with
$\rho=\rho_0=\const$, so the volume-averaged density remains constant,
i.e., $\bra{\rho}=\rho_0$.

In our simulations we change the values of magnetic and fluid
Reynolds numbers,
\EQ
\Rm=\urms/\eta\kf,\quad
\Rey=\urms/\nu\kf,
\EN
such that the ratio $\Rm/\Rey=\Pm$ has the desired value between
$10^{-3}$ and $10^3$, and we monitor the resulting kinetic and
magnetic energy dissipation rates per unit volume,
\EQ
\epsK=\bra{2\nu\rho\SSSS^2},\quad
\epsM=\bra{\eta\mu_0\JJ^2},
\label{epsdef}
\EN
whose sum, $\epsT=\epsK+\epsM$, will be used to define the
fractional dissipation rates, $\tilde\epsK=\epsK/\epsT$ and
$\tilde\epsM=\epsM/\epsT$.
We use the fully compressible {\sc Pencil Code} \footnote{
http://www.pencil-code.googlecode.com} for all our calculations.
We recall that, for the periodic boundary conditions under consideration,
$\bra{2\SSSS^2}=\bra{\WW^2}+{4\over3}\bra{(\nab\cdot\UU)^2}$,
highlighting thus the analogy between $\WW=\nab\times\UU$ and $\JJ$
in the incompressible case.

\section{Results}

In \Tab{Tsum} we summarize the parameters of runs with $\Pm$ between
$10^{-3}$ and $10^3$.
The runs with $10^{-3}\leq\Pm\leq1$ are those presented already in
Brandenburg (2009) using $512^3$ mesh points, while those with
$10\leq\Pm\leq1000$ are new ones and have been performed using $256^3$
mesh points.
In all cases, either $\Rey$ or $\Rm$ were close to the maximum possible
limit at a given resolution.
Indeed, for $\Pm=10^{-3}$ we were able to reach $\Rey=4400$ (for $512^3$
mesh points) while for $\Pm=10^3$ we could go to $\Rm=1200$ (for $256^3$
mesh points).

\begin{table}[b!]\caption{
Summary of import input and output parameters for
the runs reported in this paper.
}\vspace{12pt}\centerline{\begin{tabular}{lccccccc}
$\Pm$ & $\Rey$ & $\Rm$
& $\tilde{\epsilon}_{\rm K}$
& $\tilde{\epsilon}_{\rm M}$
& $k_{\rm K}$
& $k_{\rm M}$
& Res. \\
\hline
$10^{-3}$ & 4400&    4&  0.01&  0.99& 426&   8& $512^3$ \\
$10^{-2}$ & 2325&   23&  0.04&  0.96& 344&  25& $512^3$ \\
$10^{-1}$ & 1175&  118&  0.13&  0.87& 286&  81& $512^3$ \\
$10^{ 0}$ &  455&  455&  0.39&  0.61& 179& 201& $512^3$ \\
$10^{ 1}$ &   20&  200&  0.76&  0.24&  24&  99& $256^3$ \\
$10^{ 2}$ &    9&  850&  0.90&  0.10&  14& 263& $256^3$ \\
$10^{ 3}$ &    0&  425&  0.99&  0.01&   3& 129& $256^3$ \\
$10^{ 3}$ &    1& 1175&  0.99&  0.01&   5& 234& $256^3$ \\
\label{Tsum}\end{tabular}}\end{table}

We note that in all cases the total energy dissipation
is approximately the same.
This is perhaps not so surprising, because we keep the
amplitude of the forcing function the same.
However, the constancy of the energy dissipation rate implies that
the rate of energy injection must also be always the same and thus
independent of $\Pm$.
This means that the flow properties of the eddies at the
energy-carrying scale must be essentially independent of $\Pm$.

\begin{figure}[t!]\begin{center}
\includegraphics[width=\columnwidth]{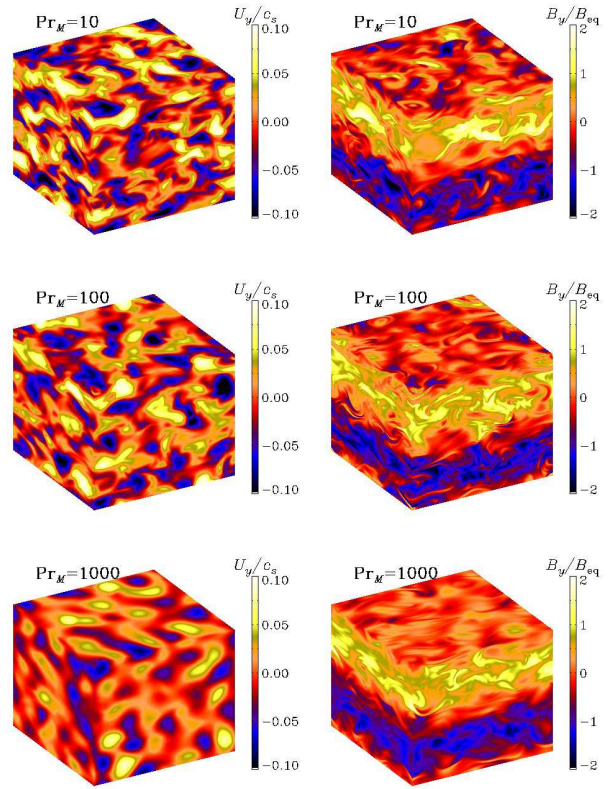}
\end{center}\caption[]{ 
Visualization of $U_y$ and $B_y$ on the periphery of the computational
domain for $\Pm$ ranging from 10 to 1000 at a resolution of $256^3$
mesh points.
}\label{UB}\end{figure}

\begin{figure}[t!]\begin{center}
\includegraphics[width=\columnwidth]{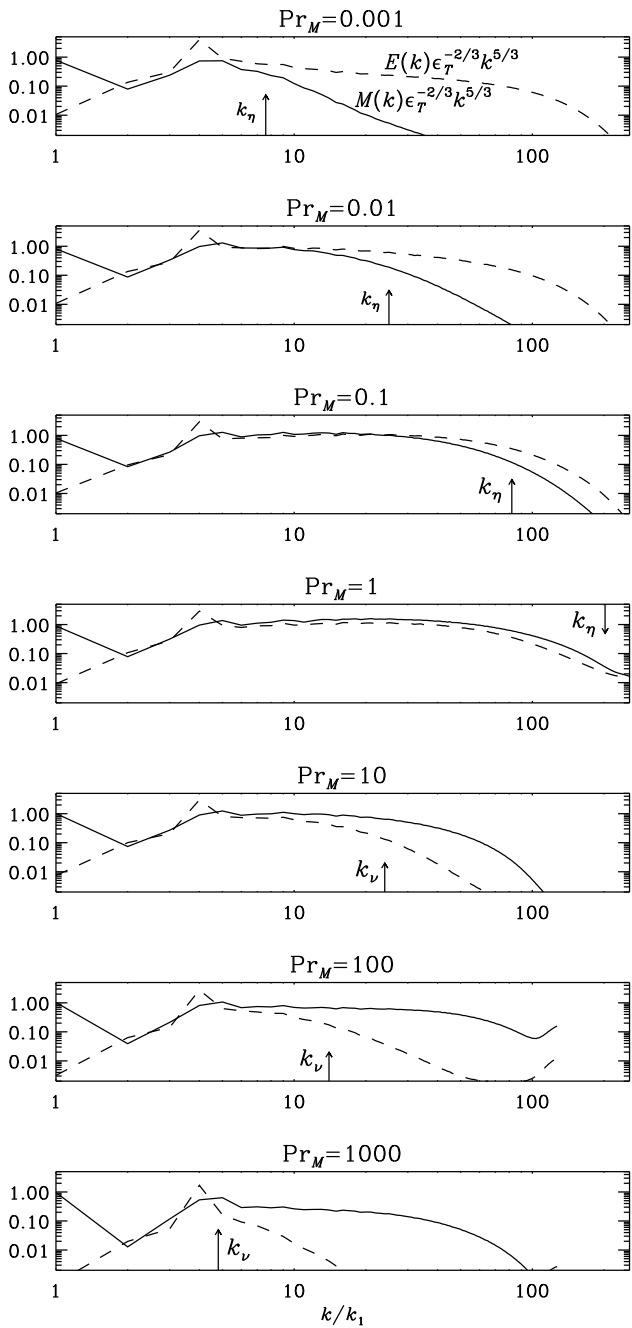}
\end{center}\caption[]{ 
Compensated kinetic and magnetic energy spectra in the saturated regime for
$\Pm=10^{-3}$ to $10^3$.
The spectra are compensated by $\epsilon_T^{-2/3}k^{5/3}$,
where $\epsilon_T$ is the sum of kinetic and magnetic energy dissipation rates.
The ohmic dissipation wavenumber, $k_\eta=(\epsilon_M/\eta^3)^{1/4}$,
is indicated by an arrow.
}\label{pspec_aver_comp_saturated2_tatry}\end{figure}

In \Fig{UB} we present visualizations of the $y$ component of velocity
and magnetic field at the periphery of the computational domain for
the new results with $\Pm\ge10$ and in 
\Fig{pspec_aver_comp_saturated2_tatry} we show spectra of
kinetic and magnetic energies, $E(k)$ and $M(k)$, respectively,
for all values of $\Pm$ between $10^{-3}$ and $10^3$.
In the velocity pattern one can clearly make out the typical
scale of the dominant eddies, whose wave length is about
1/4 of the size of the box.
The magnetic field also shows a turbulent component, but there
is a much stronger large-scale component superposed.
This is essentially the Beltrami field which is of the form
$\meanBB=(\cos k_1z,\sin k_1z, 0)$, although its wavevector could have
pointed in any of the other two coordinate directions,
$(0,\cos k_1x,\sin k_1x)$ and $(\sin k_1y, 0,\cos k_1y)$ would have been
equally probably alternatives.
We recall that all these fields are indeed the eigenfunctions of an
$\alpha^2$ dynamo problem (e.g., Brandenburg \& Subramanian 2005),
and they also emerge as the dominant field in helically driven turbulence.
It is clear that in a triply periodic domain such as that considered
here, these fields require a resistive time to reach full saturation.
For all further details we refer to Brandenburg (2001), where such a
system was studied in full detail.

Next, we consider the spectra of kinetic and magnetic energies in
\Fig{pspec_aver_comp_saturated2_tatry} which are normalized such
that $\int E(k)\,\dd k=\half\bra{\rho\UU^2}$ and
$\int M(k)\,\dd k=\half\bra{\BB^2/\mu_0}$.
It is evident from the spectra that with
increasing values of $\Pm$, the viscous dissipation wavenumber,
$k_\nu=(\epsK/\nu^3)^{1/4}$, moves to smaller and smaller values.
Analogously to the case of $\Pm\ll1$, this implies that most of the
injected energy gets dissipated by the shorter of the two cascades --
leaving only a reduced amount of energy for the other cascade.
This means that the corresponding diffusion coefficient can be
decreased further, without creating numerical difficulties.

It appears that it is not only the energy input at the small wavenumber
end of the relevant cascade  that is decreased, but that there is possibly a
continuous removal of energy along the cascade, making the spectral index
slightly steeper than $-5/3$.
For example, for $\Pm=10^{-3}$ the spectral slope of $E(k)$ is about
$-2.2$, while for $\Pm=10^3$ the spectral slope of $M(k)$ is about $-2.0$.

It is quite extraordinary that in all these cases the nature of the
large-scale dynamo is virtually unchanged, even though $\Pm$ is varied
by 6 orders of magnitude.
The reason is that in all cases the dynamo number, $C_\alpha$, exceeds
the critical value for dynamo action, $C_\alpha^{\rm crit}=1$.
Looking at \Eq{Calp}, we see that $C_\alpha$ is dominated by the
scale separation ratio, which is here $\kf/k_1\approx4$.
Furthermore, because the turbulence is nearly fully helical, we
have $\epsf\approx1$, and since $\Rm\gg1$, we have $\iota\approx1$.
Thus, we have $C_\alpha>1$ for all runs.
We recall also that the saturation amplitude of the field is essential
given by the square root of the scale separation ratio (Brandenburg 2001),
which is about 2 in units of the equipartition field strength.
This is in reasonable agreement with the simulation results; see
\Fig{pspec_aver_comp_saturated2_tatry}, where we show the resulting
spectra for all the runs.

\begin{figure}[t!]\begin{center}
\includegraphics[width=\columnwidth]{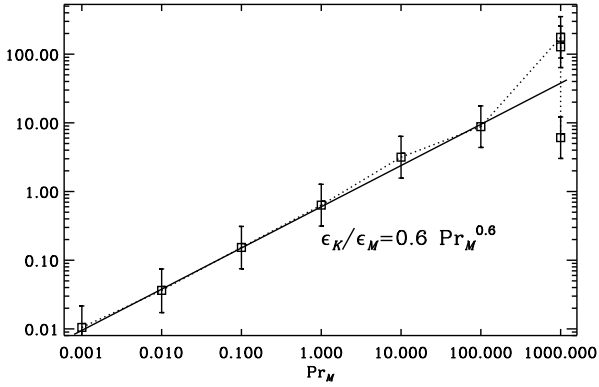}
\end{center}\caption[]{ 
Dependence of the ratio of the dissipation
rates on $\Pm$.
}\label{epsKM}\end{figure}

Next, we plot in \Fig{epsKM} the ratio of kinetic to magnetic
energy dissipation rates.
In agreement with Brandenburg (2009), we find that the ratio is
approximately proportional to $\Pm^{1/2}$, although a better fit
is now provided by $\epsK/\epsM\approx0.6\,\Pm^{\,0.6}$.
The reason for such a scaling is unclear.
However, from \Eq{epsdef} one can see that in the ratio $\epsK/\epsM$
there is an implicit proportionality with respect to $\Pm$.
Assuming, for simplicity,
$\bra{2\SSSS^2}\approx\bra{\WW^2}\approx W_{\rm rms}^2$,
we see that
\EQ
{\epsK\over\epsM}\approx\rho{\nu\over\eta}\,{W_{\rm rms}^2\over J_{\rm rms}^2}
\propto\Pm^n,
\EN
so
\EQ
{W_{\rm rms}\over J_{\rm rms}}\propto\Pm^{(n-1)/2}
\approx\Pm^{-1/4}\ldots\;\Pm^{-1/6},
\EN
where we have assumed that $n$ lies between 1/2 and 2/3,
which bracket the results seen here and in Brandenburg (2009).
These scalings are surprising in view of the usually expected
individual scalings, namely $W_{\rm rms}\propto\nu^{-1/2}$ and
$J_{\rm rms}\propto\eta^{-1/2}$ (cf.\ Brandenburg \& Subramanian 2005).

\begin{figure}[t!]\begin{center}
\includegraphics[width=\columnwidth]{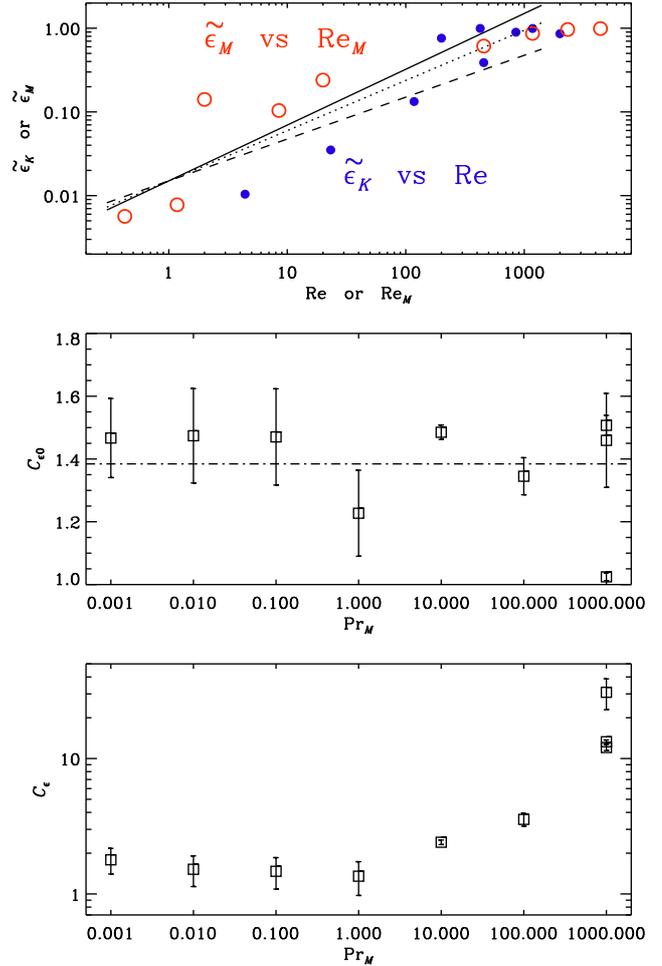}
\end{center}\caption[]{ 
{\it Top}: Dependence of $\epsK$ on $\Rey$ (blue, solid symbols) and
$\epsM$ on $\Rm$ (red, open symbols).
The solid line has the slope 2/3, while the dotted and dashed lines
have slopes 0.6 and 0.5, respectively.
{\it Middle and bottom}: scalings of $C_{\epsilon0}$ and $C_\epsilon$
versus $\Pm$.
}\label{epsKMsep}\end{figure}

In order to illuminate the issue further, we ask whether
not only the ratio $\epsK/\epsM$ scales with $\Pm$,
but whether $\epsK$ and $\epsM$ are individually proportional
to $\Rey$ and $\Rm$, respectively.
In \Fig{epsKMsep} we plot $\epsK$ versus $\Rey$ (blue, solid symbols)
and $\epsM$ versus $\Rm$ (red, open symbols).
The scatter is now much larger than in \Fig{epsKM}, and it seems that
the scaling exponent might even be as large as $n=2/3$.

We mentioned earlier that the total dissipation rate, $\epsT$,
is nearly independent of $\Pm$.
However, this is only true when we look the the dimensional
value of $\epsT$.
It is customary to consider the normalized dissipation rate,
\EQ
C_\epsilon={\epsT\over\uoneD^3/L},
\EN
where $\uoneD=\urms/\sqrt{3}$ is the one-dimensional rms velocity
and $L=3\pi/4\kf$ is conventionally used as the integral scale
(Pearson et al.\ 2004).
In the second and third panels of \Fig{epsKMsep} we compare
$C_\epsilon$ with $C_{\epsilon0}$, which is based on the
maximum value of $\uoneD$ in all the runs.
The difference is caused by the fact that $\urms$ drops to rather low
values in the large-$\Pm$ regime.
Part of this goes into magnetic energy, but it is not enough to make up for
this difference.

It is important to realize that, on average, $\epsM$ is just
the same as the rate of work done against the Lorentz force,
$-\bra{\UU\cdot(\JJ\times\BB)}$.
This becomes evident when considering the flow of energy in our system:
\begin{eqnarray}
\bra{\rho\UU\cdot\ff}\to\left\{
\begin{array}{ll}
 & \to \bra{2\rho\nu\SSSS^2}\\
-\bra{\UU\cdot(\JJ\times\BB)} & \to \bra{\eta\mu_0\JJ^2}.
\end{array}
\right.
\end{eqnarray}
Here, $\bra{\rho\UU\cdot\ff}\approx\epsT$ is the rate of energy injection
into the system by the forcing term.
Normally, in the hydrodynamic case, $\bra{2\rho\nu\SSSS^2}$,
or $\bra{\nu\WW^2}$ in the incompressible case, stay constant
as $\nu$ is decreased.
In the case with dynamo action, however, a decrease in $\nu$
allows the dynamo to tap more energy, so
$-\bra{\UU\cdot(\JJ\times\BB)}$ and $\epsM$
increase at the expense of $\epsK$.
This is indicated by the fact $\epsK/\epsM$ is found
to be proportional to $(\nu/\eta)^n$, so $\epsK$ decreases
as $\nu$ decreases.
This decease is weak in the sense that $n\approx1/2\,...\,2/3$ is
less than unity, but it is certainly no longer independent of
$\nu$ as it would be in the purely hydrodynamic case.

In view of the application to quasars, i.e.\ accretion discs in active
galactic nuclei, it is relevant to consider the fraction of energy that
goes into the heating of electrons.
Indeed, such discs are known to be underluminous, which led to the
standard paradigm of advection-dominated accretion (Narayan \& Yi 1994;
Abramowicz et al.\ 1995).
Alternatively, this might be associated with the small value of the
ratio $\epsM/\epsT$, for which we find
\EQ
{\epsM\over\epsT}={\epsM\over\epsM+\epsK}\propto{1\over1+\Pm^n}.
\EN
Using standard accretion disc theory, Balbus \& Henri (2008) find
that $\Pm$ depends on the distance $R$ from the black hole and
is proportional to $R^{-9/8}$.
In particular, they find that $\Pm$ exceeds unity within about
50 Schwarzschild radii.
This would dramatically decrease $\epsM$ in the inner parts and
might be sufficient to explain underluminous accretion.
However, this proposal hinges on several assumptions:
(i) that the viscous heating heats the ions and not the electrons,
(ii) that the resistive dissipation energizes electrons rather than ions,
(iii) that the discs are essentially collisionless and, finally,
(iv) that the magnetohydrodynamic approximation is then still applicable.

\section{Conclusions}

The present work has shown that the ratio of kinetic to magnetic
energy dissipation follows one and the same relationship with
$\Pm$ both for small and large values.
An important additional condition obeyed by all our runs is,
however, that the magnetic Reynolds number is large enough
for dynamo action to occur.
This constitutes an important difference between our current
results for large-scale dynamos and those mentioned in the
first section for small-scale dynamos.
An important consequence of such scaling is the fact that at extreme
values of the magnetic Prandtl number, larger Reynolds numbers can be
tolerated by the numerical scheme at a resolution that would be
insufficient if the magnetic Prandtl number were unity.
This was shown previously for $\Pm=10^{-3}$, in which case
fluid Reynolds numbers of 4500 were possible at a resolution
of $512^3$ meshpoints, while for $\Pm=1$ it was only possible
to reach Reynolds numbers of less that 700.
Both cases obeyed the empirical constraint that the spectral {\it kinetic}
energy has developed a clear dissipative subrange with an exponential
decay shortly before the Nyquist frequency.  In the opposite case of
large $\Pm$, here $\Pm=10^3$, it was possible to reach magnetic Reynolds
numbers of 1000 at $256^3$ mesh points.
In this case the {\it magnetic} energy spectrum has developed a
dissipative subrange shortly before the Nyquist frequency,
although it was less convincing for $\Pm=10^2$.

The reason for the value of the exponent $n$ in the power law relation
between the energy dissipation ratio $\epsK/\epsM$ and $\Pm$ remains unclear.
At this point we cannot be certain that it is $n=0.6$ and not, for
example,  1/2 or 2/3.
One source of error might come from the fact that at extreme values of
$\Pm$ the effects of numerical viscosity associated with the advection
operator are no longer negligible.
For the third-order time step used in the {\sc Pencil Code}, the
numerical viscosity operator takes the form $-\nu_2^{\rm CFL}\nabla^4$
where $\nu_2^{\rm CFL}=\urms\delta x^3C_{\rm CFL}^3/24$ is a numerical
hyperviscosity\footnote{See page 118 of the {\sc Pencil Code} manual,
\url{http://www.nordita.org/software/pencil-code/doc/manual.pdf}}
that depends on the mesh size $\delta x$ and the
Courant--Friedrich--Levy number $C_{\rm CFL}$, whose default value is 0.4,
but the code would still be numerically stable for $C_{\rm CFL}=0.9$.
If such numerical effects do begin to play a role, we must expect that
the effective values of $\Pm$ are less extreme, which means that the $n$
would have been underestimated and that $n$ might be 2/3 or even larger.

While earlier work focussed on the dependence of $\epsM$ on $\Pm$
(Blackman \& Field 2008), no clear conclusion about the dissipation
ratio $\epsK/\epsM$ seems to have emerged.
For example, if $\epsK$ and $\epsM$ were independent of viscosity and
magnetic diffusivity, the ratio $\epsK/\epsM$ would have been constant.
Instead, we find that $\epsK$ decreases when $\Rey$ decreases,
and likewise, $\epsM$ decreases when $\Rm$ decreases.
On the other hand, one must be cautious when applying results
regarding the dependence on $\Rm/\Rey$ ($=\Pm$) for large values
of $\Rey$ and $\Rm$, because we may still not be in an asymptotic
parameter regime.
It is therefore important to extend this work to larger values of
$\Rey$ and $\Rm$ and to go to larger numerical resolution.

\acknowledgements
I thank the referee for making several useful suggestions.
The computations have been carried out on the
National Supercomputer Centre in Link\"oping and the Center for
Parallel Computers at the Royal Institute of Technology in Sweden.
This work was supported in part by the Swedish Research Council,
grant 621-2007-4064, and the European Research Council under the
AstroDyn Research Project 227952.


\end{document}